\documentstyle[aps,preprint]{revtex}
\tightenlines
\begin{document}
\draft
\title{"quasi-particles" in bosonization theory of interacting fermion liquids
at arbitrary dimensions}
\author{Tai-Kai Ng}
\address{
Department of Physics,
Hong Kong University of Science and Technology,
Clear Water Bay Road,
Kowloon, Hong Kong
}
\date{ \today }
\maketitle
\begin{abstract}
   Within bosonization theory we introduce in this paper a new definition 
of "quasi-particles" for interacting fermions at arbitrary space dimenions. 
In dimensions higher than one we show that the constructed quasi-particles 
are consistent with quasi-particle descriptions in Landau Fermi liquid 
theory whereas in one-dimension the "quasi-particles" are non-perturbative
objects (spinons and holons) obeying fractional statistics. The 
more general situation of Fermi liquids with singular Landau interaction 
is discussed.

\end{abstract}


\narrowtext

  The concept of "quasi-particles", as excitations adiabatically connected
to excitations in corresponding non-interacting Fermi gas has formed the 
basis for our understanding of Fermi liquids at dimensions higher than 
one\cite{t1}. Based on Bethe-ansatz solutions\cite{t2} of exactly 
solvable models, it is now understood that similar to fermions in higher 
dimensions, the low energy behaviours of interacting fermions in one 
dimension can also be described as gases of "free", or quasi-particles, 
except that these particles are not adiabatically connected to free 
fermions as in Landau fermi-liquid theory, but are non-perturbative 
objects called "spinon" and "holon" characterizing the spin- and 
charge- degrees of freedom of the system separately\cite{t3}. The spinons 
and holons are neither fermions or bosons, but are objects obeying
fractional(exclusion) statistics\cite{t4}.

  The purpose of the present paper is to show that the two rather 
different pictures of "quasi-particles" in Landau Fermi liquid and in
Bethe-ansatz solutions in one dimension, can be understood in 
bosonization theory by introducing a new definition of "quasi-particle" 
operators. Our approach provides a simple way of unifying the two
"quasi-particle" pictures and provides a more general framework of 
understanding quasi-particles in interacting fermion systems. We shall 
also discuss the situation of generalized Fermi liquids with singular 
Landau interaction within our framework.

  To warm up we first review briefly bosonization theory in one\cite{t5}
and higher dimensions\cite{t6,t7,t8}. The main idea of bosonization theory 
is that the low energy excitation spectrum of Fermi and Luttinger liquids 
can both be described by effective harmonic theories describing elastic
deformation of Fermi surfaces\cite{t7,t8}. The deformation at position
$\vec{k}_F$ on the Fermi surface is described by the coarse-grained Wigner operator\cite{t8}
\[
\rho_{\vec{k}_F\sigma}(\vec{q})={1\over{V}}\sum_{-\Lambda<k<\Lambda}
c^+_{\vec{k}_F+k\hat{k}_F+\vec{q}/2\sigma}c_{\vec{k}_F+k\hat{k}_F-
\vec{q}/2\sigma}, \]
where $V=$volume, $\hat{k}_F$ is a unit vector along the direction of
$\vec{k}_F$ and $\Lambda<<k_F$ is a high momentum cutoff. $c(c^+)$ are
the usual fermion annihilation(creation) operators. We shall set $\hbar=1$
in this paper. Notice that in one dimension 
$\vec{k}_F\rightarrow\pm{k}_F$ and $\rho_{\vec{k}_F\sigma}(\vec{q})
\rightarrow\rho_{L(R)\sigma}(q)$, i.e. left and right Fermi points. 
$\rho_{\vec{k}_F\sigma}(\vec{q})$'s obey approximate commutation 
relations\cite{t7,t8}
\begin{equation}
\label{com}
[\rho_{\vec{k}_F\sigma}(\vec{q}),\rho_{\vec{k}'_F\sigma'}(-\vec{q}')]
=-\delta^D(\vec{q}-\vec{q}')\delta^{D-1}(\vec{k}_F-\vec{k}'_F)
\delta_{\sigma\sigma'}{\vec{q}.\vec{k}_F\over{m}}N_{\Lambda}(0),
\end{equation}
where $D$=dimension and $N_{\Lambda}(0)=N(0)/S_d$, where $N(0)$ is the 
density of states of spin-$\sigma$ fermions on Fermi surface, and
$S_d=\int{d}\Omega=$ solid angle on the Fermi surface. For computation 
purpose it is convenient to introduce canonical boson operators 
$b^+(b)$ defined by $\rho_{\vec{k}_F\sigma}(\vec{q})=
\left({|\vec{k}_F.\vec{q}|\over{m}}N_{\Lambda}(0)\right)^{1\over2}
(\theta(\vec{k}_F.\vec{q})b^+_{\vec{k}_F\sigma}(\vec{q})+
\theta(-\vec{k}_F.\vec{q})b_{\vec{k}_F\sigma}(-\vec{q}))$.
The $b(b^+)$ operators satisfy canonical boson commutation relations\cite{t8}.

  The low energy physics of the systems is described by effective 
Hamiltonian $H=H_0+H_1$, where $H_0$ is the kinetic energy and $H_1$ is 
the interaction. In terms of $b(b^+)$ operators, $H_0$ takes the form\cite{t8}
\begin{equation}
\label{H0}
H_0={1\over{V}}\sum_{\vec{k}_F,\vec{q},\sigma}{|\vec{k}_F.\vec{q}|\over{m}}
b^+_{\vec{k}_F\sigma}(\vec{q})b_{\vec{k}_F\sigma}(\vec{q}),
\end{equation}
where we have linearized the fermion dispersion around the Fermi surface
to obtain \ (\ref{H0}). $H_1$ can be written as
\begin{equation}
\label{H1}
H_1={1\over2V}\sum_{\vec{k}_F,\sigma,\vec{k}'_F,\sigma',|\vec{q}|<\Lambda}
f_{\vec{k}_F\sigma\vec{k}'_F\sigma'}(\vec{q})\rho_{\vec{k}_F\sigma}(\vec{q})
\rho_{\vec{k}'_F\sigma'}(-\vec{q}),
\end{equation}
where $f_{\vec{k}_F\sigma\vec{k}'_F\sigma'}(\vec{q})$'s are effective 
parameters characterizing the residual (marginal) fermion-fermion 
interactions in the low energy and small wavevector limit\cite{t8,t9}.  
With Eqs. \ (\ref{com}) to \ (\ref{H1}), the Heisenberg equation of motion 
for $\rho_{\vec{k}_F\sigma}(\vec{q})$ is\cite{t8}
\begin{equation}
\label{landau}
\left(i{\partial\over\partial{t}}+{\vec{k}_F.\vec{q}\over{m}}\right)
\rho_{\vec{k}_F\sigma}(\vec{q})=-{\vec{k}_F.\vec{q}\over{m}}
N_{\Lambda}(0)\sum_{\vec{k}'_F\sigma'}f_{\vec{k}_F\sigma\vec{k}'_F\sigma'}
(\vec{q})\rho_{\vec{k}'_F\sigma'}(\vec{q}),
\end{equation}
and is the same as Landau transport equation for quasi-particles in 
Fermi liquid theory in the $\vec{q}\rightarrow0$ limit where
$f_{\vec{k}_F\sigma\vec{k}'_F\sigma'}(\vec{q}\rightarrow0)$ can be
identified as Landau parameters\cite{t8,t9}. The same form of equation 
of motion for $\rho_{L(R)\sigma}(q)$ is also obtained at one 
dimension. 

  Although the equations of motion have the same form at one and higher 
dimensions, the resulting eigen-state spectrums are very different. At
higher dimensions there are two kinds of eigen-state solutions to 
Eq.\ (\ref{landau}), the particle-hole continuum and collective 
modes\cite{t10}. The particle-hole continuum are eigenstates 
adiabatically connected to the particle-hole spectrum of 
non-interacting fermions, whereas collective modes are non-perturbative
eigenstates that disappear in the absence of interaction. The validity 
of Fermi liquid theory is warranted by the existence of adiabatic 
particle-hole spectrum\cite{t10}. In one dimension {\em only} 
non-perturbative collective density and spin wave excitations 
exist\cite{t5}. The collective excitations are not adiabatically 
connected to excitation spectrum of non-interacting fermions showing 
that the systems are not Landau Fermi liquids. 

  Next we consider the construction of quasi-particle operators. Our goal 
is to search for a bosonization representation of the quasi-particle 
operator that is valid in the low energy, long wave length limit. To 
gain insight we first consider non-interacting fermions. In the 
spirit of Landau Fermi liquid theory we consider a (spin-$\sigma$)
quasi-particle wavepacket with momentum $\vec{k}_F$ at position $\vec{r}$. 
Notice that quantum mechanics requires that there exist uncertainties
$\delta{r}\sim\Lambda^{-1}$ and $\delta{k}\sim\Lambda$ in the position 
and momentum of the particle. For non-interacting fermions the 
quasi-particle wavepacket can be identified as a free 
spin-$\sigma$ fermion wavepacket described by
\[
\psi_{\vec{k}_F\sigma}(\vec{r})\sim{1\over2\pi}\int^{\Lambda}_{-\Lambda}
k^{D-1}dke^{-ik\hat{k}_F.\vec{r}}c_{\vec{k}_F+k\hat{k}_F\sigma},  \]
where we have assumed that there is no uncertainty in the {\em direction}
$\hat{k}_F$. In the limit $\Lambda<<k_F$ it is easy to see that the 
wave-packet operator satisfies the equation of motion 
\begin{equation}
\label{em1}
i{\partial\over\partial{t}}\psi_{\vec{k}_F\sigma}(\vec{r})
=[\psi_{\vec{k}_F\sigma}(\vec{r}),H_o-\mu{N}]={i\vec{k}_F.\nabla\over{m}}
\psi_{\vec{k}_F\sigma}(\vec{r})+O(\Lambda/k_F).
\end{equation}
where $N=$ number of particles and $\mu$ is the chemical potential.

   We shall now show that we can use Eq.\ (\ref{em1}), which specifies 
the dynamics of a free-particle wavepacket, as the {\em definition} of  quasi-particle operators for free fermions in bosonization theory.
To see this we write $\psi_{\vec{k}_F\sigma}(\vec{r})=
\hat{f}_{\hat{k}_F}e^{J_{\vec{k}_F\sigma}(\vec{r})}$,
where $J_{\vec{k}_F\sigma}(\vec{r})\sim\sum_{\vec{q}}\alpha_{\vec{k}_F\sigma}
(\vec{q},\vec{r})\rho_{\vec{k}_F\sigma}(\vec{q})$ is a linear combination 
of $\rho_{\vec{k}_F\sigma}(\vec{q})$ operators, i.e. $\psi_{\vec{k}_F\sigma} 
(\vec{r})$ represents a coherent state of bosonic waves in bosonization 
theory. $\hat{f}_{\hat{k}_F}$ is an operator (Klein factor) introduced 
to ensure the anti-communication relation between $\psi_{\vec{k}_F\sigma}
(\vec{r})$ fields in different directions $\vec{k}_F$'s\cite{t6}. The linear 
coefficients $\alpha$'s characterizing $J_{\vec{k}_F\sigma}(\vec{r})$ 
are determined by requiring that $\psi_{\vec{k}_F\sigma}(\vec{r})$ 
satisfies the equation of motion \ (\ref{em1}) with the bosonized 
$H_0$ (Eq. \ (\ref{H0})). After some straightforward algebra, we obtain
\begin{equation}
\label{j}
J_{\vec{k}_F\sigma}(\vec{r})\sim-(e^*){1\over{V}}\sum_{\vec{q}}
{me^{-i\vec{q}.\vec{r}}\over{N}_{\Lambda}(0)\vec{k}_F.\vec{q}}
\rho_{\vec{k}_F\sigma}(\vec{q}),
\end{equation}
where $e^*$ is an arbitrary number that is not fixed by the equation
of motion. The meaning of $e^*$ can be seen by examining
the communication relation between $\psi$ operator and the total charge
operator $\rho(\vec{q})=\sum_{\vec{k}_F,\sigma}\rho_{\vec{k}_F\sigma}
(\vec{q})$. We obtain
\[
[\psi_{\vec{k}_F\sigma}(\vec{r}),\rho(-\vec{q})]
=(e^*)e^{-i\vec{q}.\vec{r}}\psi_{\vec{k}_F\sigma}(\vec{r}),  \]
showing that $e^*$ represents the charge carried by the quasi-particle 
and is equal to one for free fermions. The reason why we cannot determine 
$e^*$ by the equation of motion is that the solutions of 
Eq. \ (\ref{em1}) represent coherent superposition of boson waves 
travelling with constant velocity $\vec{v}_F=\vec{k}_F/m$, and
superposition of different solutions is again a solution to the equation. 
We note that in usual bosonization theory $\psi_{\sigma}(\vec{r}) 
=\sum_{\vec{k}_F}\psi_{\vec{k}_F\sigma}(\vec{r})$ is identified as the
bosonization representation of fermion operator and is usually derived 
by requiring that $\psi_{\sigma}(\vec{r})$ satisfies the correct 
commutation relation with the density operator\cite{t6,t8}. We define
$\psi_{\sigma}(\vec{r})$ as the quasi-particle operator through the
equation of motion in our approach. For free fermions the two 
definitions give identical result. 

  In the presence of interaction we define quasi-particle operators as 
operators representing coherent states of bosonic waves satisfying 
equations of motion of form
\begin{equation}
\label{em}
i{\partial\over\partial{t}}\psi_{\vec{k}_F\gamma}^{(Q)}(\vec{r})
=[\psi_{\vec{k}_F\gamma}^{(Q)}(\vec{r}),H-\mu{N}]\sim{i}\vec{v}_{\gamma}.\nabla
\psi_{\vec{k}_F\gamma}^{(Q)}(\vec{r})+O(\Lambda/k_F).
\end{equation}
where $H$ is the full bosonized Hamiltonian and $\gamma$ is a branch index. 
$\vec{v}_{\gamma}\sim\vec{k}_F/m_{\gamma}$ is the velocity of branch 
$\gamma$ quasi-particles at position $\vec{k}_F$ on the Fermi surface. 
The nature of the different branches and corresponding $\vec{v}_{\gamma}$ 
are determined self-consistently by requiring that
$\psi_{\vec{k}_F\gamma}^{(Q)}(\vec{r})$ satisfies the equation of 
motion \ (\ref{em}).
Writing $\psi_{\vec{k}_F\gamma}^{(Q)}(\vec{r})=\hat{f}_{\hat{k}_F}
e^{J^{(Q)}_{\vec{k}_F}(\gamma,\vec{r})}$ where $J_{\vec{k}_F\sigma}(\vec{r})
\sim\sum_{\vec{q}}\alpha_{\vec{k}_F\sigma}(\vec{q},\vec{r})
\rho_{\vec{k}_F\sigma}(\vec{q})$ as before, we obtain from \ (\ref{em})
a linear eigenvalue equation for the coefficents determining $\alpha$. 
The nature of the quasi-particle branches and $\vec{v}_{\gamma}$ are 
determined by the eigenvectors and eigenvalues of the eigenvalue 
equation. 

  To illustrate we consider interacting fermions at dimensions $D>1$. 
A easy way to obtain the quasi-particle operators is to notice that 
at dimenions higher than one, a continuous spectrum of
particle-hole pair solution $\bar{\rho}_{\vec{p}_Fs}(\vec{q})=
\rho_{\vec{p}_Fs}(\vec{q})+\sum_{\vec{k}_F\sigma}\xi_{\vec{p}_Fs
\vec{k}_F\sigma}(\vec{q})\rho_{\vec{k}_F\sigma}(\vec{q})$ to Eq. \
(\ref{landau}) exists, with eigen-energy $\epsilon=\vec{p}_F.\vec{q}/m$ and 
\[
\xi_{\vec{p}_Fs\vec{k}_F\sigma}(\vec{q})={1\over{V}}\left({{\vec{p}_F.\vec{q}\over{m}}
\over{{\vec{k}_F.\vec{q}\over{m}}-{\vec{p}_F.\vec{q}\over{m}}}}\right)
N_{\Lambda}(0)A_{\vec{k}_F\sigma\vec{p}_Fs}(\vec{q}),  \]
where $A_{\vec{k}_F\sigma\vec{p}_Fs}(\vec{q})$ is the quasi-particle 
scattering matrix in Landau Fermi Liquid theory\cite{t8,t10}. A 
quasi-particle operator with momentum $\vec{p}_F$, $\psi_{\vec{p}_Fs}^{(Q)}
(\vec{r})=\hat{f}_{\hat{p}_F}e^{J^{(Q)}_{\vec{p}_F}(s,\vec{r})}$ 
satisfying equation of motion \ (\ref{em}) can be obtained by choosing 
\begin{equation}
\label{jq}
J^{(Q)}_{\vec{p}_F}(s,\vec{r})=-{1\over{V}}\sum_{\vec{q}}
{me^{-i\vec{q}.\vec{r}}\over{N}_{\Lambda}(0)\vec{p}_F.\vec{q}}
\bar{\rho}_{\vec{p}_F\sigma}(\vec{q}).
\end{equation}
  The branch index is given by $\gamma=s=$ spin index and $m_{\gamma}=m$ as 
for non-interacting fermions. The quasi-particle operators constructed this 
way represent bare-fermions dressed by particle-hole pair excitations 
(described by the coefficients $\xi$) and are adiabatically connected to 
the non-interacting fermions. Notice that we have set $e^*=1$ to ensure
adiabaticity. We expect that the quasi-particle operator constructed 
this way corresponds to the eigen-quasi-particle state in Fermi liquid 
theory\cite{t10}. To confirm we compute the charge carried by the 
quasi-particle operator we constructed. We obtain
\begin{equation}
\label{qcharge}
[\psi_{\vec{p}_Fs}^Q(\vec{r}),\rho(-\vec{q})]
=\left(1-\sum_{\vec{k}_F\sigma}\xi_{\vec{k}_F\sigma\vec{p}_Fs}(\vec{q})\right)   e^{-i\vec{q}.\vec{r}}\psi_{\vec{p}_Fs}^Q(\vec{r}),
\end{equation}
suggesting that the charge $<\rho(\vec{q})>$ carried by the quasi-particle
excitation is $<\rho(\vec{q})>=1-\sum_{\vec{k}_F\sigma}\xi_{\vec{k}_F\sigma\vec{p}_Fs}
(\vec{q})$, in exact agreement with Fermi liquid theory where the $\xi$
factors describe screening effect\cite{t10}. The same agreement with 
Fermi liquid theory is also obtained for the current and spin operators.

  Next we consider fermions in one dimension. We consider bosonized
Hamiltonian $H$ with $\vec{k}_F\rightarrow\pm{k}_F=R,L$ and
$f_{\vec{k}_F\sigma\vec{k}'_F\sigma'}(\vec{q})=f_s(q)+
f_a(q)\vec{\sigma}.\vec{\sigma}'$, i.e., we neglect current-current
interactions. The resulting equation of motion \ (\ref{landau}) for
bosonic excitations has two branches of solution with dispersions 
$\epsilon_{s(a)}(q)=(\sqrt{v_F^2+2f_{s(a)}(q)v_F/\pi})q=v_{s(a)}q$, where 
$v_F=k_F/m$ and $s,a$ represents density and spin wave fluctuations, 
respectively\cite{t5}. Notice that $v_s\neq{v}_a$ as long as
$f_s(q)\neq{f}_a(q)$, reflecting general spin-charge separation of 
interacting fermions at one dimension.

  To construct quasi-particle operators we proceed as before and
define quasi-particle operators using Eq.\ (\ref{em}), with
$\psi_{\vec{p}_F\gamma}^{(Q)}(\vec{r})\rightarrow\psi_{L(R)\gamma}^{(Q)}
(x)$, and $(\vec{k}_F.\nabla)/m\rightarrow\pm{v}_F\partial/\partial{x}$.
Writing $\psi_{L(R)\gamma}^{(Q)}(x)=\hat{f}_{L(R)}e^{J^{(Q)}_{L(R)}
(\gamma,x)}$ where $J^{(Q)}$ is linear in the $\rho_{L(R)\sigma}(\vec{q})$ 
operators, we obtain from solving Eq.\ (\ref{em}),
\begin{equation}
\label{j1d}
J^{(Q)}_{R(L)}(\gamma,x)={e^*_{\gamma}\over2V}\sum_{q}{e^{-iqx}\over{q}}
\left(({+(-)1\over\eta_{\gamma}})(\rho_{L\gamma}(q)+\rho_{R\gamma}(q))+
(\rho_{R\gamma}(q)-\rho_{L\gamma}(q))
\right),
\end{equation}
where $\gamma=s,a$ and $\rho_{L(R)[s(a)]}(q)=\sum_{\sigma}
\left[1(\sigma)\right]\rho_{L(R)\sigma}(q)$ is the coarse-grained 
density(spin) Wigner operator. Notice that $\psi^{(Q)}_{L(R)s(a)}(x)$'s
constructed this way represent "quasi-particles" corresponding to 
coherent states formed by density (s) and spin (a) bosonic waves
separately (spin-charge separation) and are not adiabatically connected 
to the original fermions. The self-determined quasi-particle velocities
$v_{s(a)}$ are equal to the corresponding boson density(spin) wave 
velocities and $\eta_{s(a)}=v_F/v_{s(a)}$. The magnitude of charge(spin)
carried by the quasi-particle $e^*_{s(a)}$ can be determined if we 
identify the quasi-particle excitations we constructed as holons 
($\gamma=s$) and spinons ($\gamma=a$) in Bethe Ansatz solutions. The
charge(spin) carried by the holon(spinon) equals 1(1/2) and 
$e^*_s=e^*_a=1$ correspondingly.

  The quasi-particles we constructed carry fractional (excluson) 
statistics. The statistics of the quasi-particles can be determined 
directly by examining the commutation rules between quasi-particles. 
We find that the spin and charge quasi-particles satisfy non-trivial 
commutation relation among themselves. We obtain 
\[
\psi_{L(R)\gamma}^{(Q)}(x)\psi_{L(R)\gamma}^{(Q)}(x')
=e^{(i\pi/\eta_{\gamma})sgn(x-x')}\psi^{(Q)}_{L(R)\gamma}(x')
\psi^{(Q)}_{L(R)\gamma}(x), \]
indicating that the charge and spin quasi-particles are exclusons with 
statistical parameter $1/\eta_s$ and $1/\eta_a$, respectively, in
agreement with previous results\cite{t4,t11}.

  A natural question that arise is what is the most general criteria
that quasi-particle operators can be constructed in a fermionic system
at dimensions $>1$, given that the low energy dynamics of the system 
is described by equation of motion of form \ (\ref{landau})? This question 
is important because it has been observed in recent years that non-Fermi 
liquid behavior appears in certain systems at two dimension, for 
example, fermions in half-filled Landau level or in strongly fluctuating 
gauge fields\cite{t12,t13}. The low energy physics of these systems are 
believed to be described by transport equation similar to 
Eq.\ (\ref{landau}), except that the effective Landau interactions and
corresponding effective masses are frequency dependent and are singular 
in the $\omega,q\rightarrow0$ limit\cite{t12,t13}. 

  For frequency-independent Landau interactions it can be shown that
quasi-particle operators defined by Eq.\ (\ref{em}) can be constructed
as long as there exists a continuum eigen-spectrum of the equation of 
motion \ (\ref{landau}) with eigenenergy $\epsilon_{\vec{p}_F}(\vec{q})
=\vec{v}_{\gamma}.\vec{q}$ at small q, at every point $\vec{p}_F$ on 
the Fermi surface. The solutions may or may not be adiabatically
connected to the non-interacting fermions. The physical reason for this
requirement is clear: if the low energy excitations of the system is
characterized by quasi-particle occupation numbers with quasi-particle
dispersion that is continuous across the Fermi surface, then 
particle-hole spectrum of the form $\epsilon_{\vec{p}_F}(\vec{q})
=\vec{v}_{\gamma}.\vec{q}$ must exist for all $\vec{p}_F$ at small $q$. 
What is surprising is that this is also a {\em sufficient} condition for 
existence of quasi-particles in bosonization theory with our definition. 
 
  For transport equations with frequency-dependent Landau parameters
the above conclusion does not hold because the eigenstates of the 
transport equations are not orthogonal to each other, although we expect 
that the conclusion may be qualitatively similar. This seems to be the
case for fermions in half-filled Landau level or in fluctuations transverse 
gauge fields\cite{t12}. Notice that in the case when the particle-hole 
continuum solution is non-perturbative, we expect that some Ward
identities derived for Landau Fermi liquids may be violated, and 
non-Fermi liquid behaviours may appear in the physical response functions.

  The author thanks Prof. Y. B. Kim for helpful comments.


\begin{references}
\bibitem{t1} L.D. Landau, Sov. Phys. JEPT, {\bf 3}, 920 (1956), {\em ibid},
 {\bf 8}, 70 (1959). 
\bibitem{t2} See, e.g., B. Sutherland, in {\em Exactly Solvable Problems in
  Condensed Matter and Relativistic Field Theory}, edited by B.S. Shastry,
  S.S. Jha, and V.Singh, Lecture Notes in Physics Vol. 242 (Springer,
  Berlin, 1985), P.1.
\bibitem{t3} M. Ogata and H. Shiba, \prb {\bf 41}, 2326 (1990).
\bibitem{t4} F.D.M. Haldane, \prl {\bf 67}, 837 (1991); Y. Hatsugai,
 M.Kohmoto, T. Koma and Y.S. Wu, \prb {\bf 54}, 5328 (1996).
\bibitem{t5} S. Tomonaga, Prog. Theor. Phys. {\bf 5}, 544(1950); 
 F.D.M. Haldane, J. Phys. C {\bf 14}, 2585(1981).
\bibitem{t6} A. Luther, \prb {\bf 19}, 320 (1979). 
\bibitem{t7} F.D.M. Haldane, in {\em Luttinger's Theorem and Bosonization
 of the Fermi Surface}, Proceedings of the Interbational School of 
 Physics "Enrico Fermi", Course CXXI, Varenna, 1992, edited by R. Schrieffer
 and R.A. Broglia (North-Holland, NY 1994).
\bibitem{t8} A. Houghton and J.B. Marston, \prb {\bf 48}, 7790(1993);
  A.H. Castro Neto and E. Fradkin, \prl {\bf 72}, 1393(1994).
\bibitem{t9} R. Shankar, Rev. Modern Phys. {\bf 66}, 129(1994).
\bibitem{t10} P.Nozieres and D. Pines, in {\em The Theory of Quantum
 Liquids}, Advanced Book Classics, edited by D. Pines (Peresus Books,
 Cambridge 1996).
\bibitem{t11} Y.S. Wu and Y.Yu, \prl {\bf 75}, 890 (1995).
\bibitem{t12} Y.B. Kim, P.A. Lee and X.G. Wen, \prb {\bf 52}, 17275 (1995);
  A. Stern and B.I. Halperin, \prb {\bf 52}, 5890 (1995).
\bibitem{t13} P.A. Bare and X.G. Wen, \prb {\bf 48}, 8636(1993); Y.L. Liu
  and T.K.Ng, \prl {\bf 83}, 5539 (1999).
\end{references}
\end{document}